\documentclass[oldversion]{aa}  
\usepackage{graphicx,amsmath}
\usepackage{wasysym}
\usepackage{rotating}
\usepackage{color}
\usepackage{natbib}             
\usepackage{url}
\usepackage{multirow}
\bibpunct{(}{)}{;}{a}{}{,} 

\begin{document}

   \title{The Mg\,{\sc i} line: a new probe of the atmospheres of evaporating exoplanets.}
                                   
   \author{
   V.~Bourrier\inst{1,2,3}\and
   A.~Lecavelier des Etangs\inst{1,2}\and
   A.~Vidal-Madjar\inst{1,2}
        }
   
\authorrunning{V.~Bourrier et al.}
\titlerunning{Survey of escaping atmospheres using the Mg\,{\sc i} line}

\offprints{V.B. (\email{vincent.bourrier@unige.ch})}

   \institute{
   CNRS, UMR 7095, 
   Institut d'astrophysique de Paris, 
   98$^{\rm bis}$ boulevard Arago, F-75014 Paris, France
   \and
   UPMC Univ. Paris 6, UMR 7095, 
   Institut d'Astrophysique de Paris, 
   98$^{\rm bis}$ boulevard Arago, F-75014 Paris, France
   \and 
	 Observatoire de l'Universit\'e de Gen\`eve, 51 chemin des Maillettes, 1290 Sauverny, Switzerland
	 }
   
   \date{} 
 
  \abstract
{

Transit observations of HD\,209458b in the UV revealed signatures of neutral magnesium escaping the planet's upper atmosphere. The absorption detected in the Mg\,{\sc i} line provides unprecedented information on the physical conditions at the altitude where the atmospheric blow-off takes place. Here we use a 3D model of atmospheric escape to estimate the transit absorption signatures in the Mg\,{\sc i} line of their host stars. The detectability of these signatures depends on the brightness of the star and the escape rate of neutral magnesium. We identify a sample of potentially evaporating exoplanets that covers a wide range of stellar and planetary properties, and whose extended exospheres might be detected through Mg\,{\sc i} line observations with current UV facilities, allowing further steps in comparative exoplanetology.

%
}


   \maketitle

\section{Introduction}
\label{intro} 

Planetary atmospheres outside our solar system have been characterized using observations of transiting planets. In particular, Lyman-$\alpha$ observations of atomic hydrogen in the atmospheres of HD\,209458b and HD\,189733b revealed that hot Jupiters "evaporate", significant amounts of gas escaping their gaseous atmosphere (e.g., \citealt{VM2003}; \citealt{Lecav2010}). Heavier elements were identified at high altitudes in the extended exosphere of HD\,209458b, in the lines of O\,{\sc i}, C\,{\sc ii,} and Si\,{\sc iii} (\citealt{VM2004}; \citealt{Linsky2010}, \citealt{BJ_Hosseini2010}), Si\,{\sc iv} (\citealt{Schlawin2010}), and Mg\,{\sc i} (\citealt{VM2013}). Signatures of escaping particles have also been found in the Lyman-$\alpha$ line of the warm Jupiter 55\,Cnc b (\citealt{Ehrenreich2012}) and the hot-Neptune GJ436b (\citealt{Kulow2014}), and in the Mg\,{\sc ii} line of the hot Jupiter WASP-12b (\citealt{Fossati2010}; \citealt{Haswell2012}). A wide variety of models has been developed to explain these observations, which are well explained by the "blow-off" of the atmosphere (see \citealt{Bourrier_lecav2013} and references therein): the strong X/EUV heating from the star heats the whole upper atmosphere to such a degree that it is placed in a hydrodynamic state, in some cases overflowing the Roche lobe (\citealt{Lecav2004}). Observations of escaping atmospheres are for now limited to very few cases, and yet estimates of the evaporation state of transiting exoplanets from determinations
of their ``energy budget'' (e.g., \citealt{Lecav2007}; \citealt{Ehrenreich_desert2011}) show that smaller planets may also bear significant mass losses, as shown by the recent result of \citet{Kulow2014} for the hot Neptune GJ\,436 b. Here we propose to evaluate the evaporation state of known transiting exoplanets and to calculate the dynamics of escaping neutral magnesium atoms in order to estimate their detectability with current facilities.


\section{Atmospheric escape of neutral magnesium from HD\,209458b}
\label{obs}

Using the echelle E230M grating of the Space Telescope Imaging Spectrograph (STIS) instrument onboard the Hubble Space Telescope (HST), \citet{VM2013} observed the transit of the hot Jupiter HD\,209458b in the lines of neutral (2852.9641\,\AA) and singly ionized magnesium (2796.3518 and 2803.5305\,\AA). While no atmospheric absorption was observed in the Mg\,{\sc ii} line, the authors detected a blue-shifted transit absorption signature in the velocity range -62 to -19\,km\,s$^{−1}$ in
the Mg\,{\sc i} line, with an absorption depth of 8.8$\pm$2.1\%. \citet{Bourrier2014} used a 3D particle model of the dynamics of the escaping magnesium atoms, coupled with an analytical modeling of the atmosphere below the exobase, to reproduce these observations. They found that the observations are best reproduced if the exobase is close to the Roche lobe and neutral magnesium particles escape from the exobase at a rate of 2.9$\stackrel{+0.5}{_{-0.9}}\times10^{7}$\,g\,s$^{-1}$ with a velocity of 25\,km\,s$^{-1}$.


\section{Modeling magnesium escape from transiting exoplanets}

To identify the best targets for observations of magnesium escape, we applied a modified version of the model of \citet{Bourrier2014} to the known transiting exoplanets. We calculated the theoretical absorption profile generated by neutral magnesium in their atmosphere to derive the significance level of its signature in the Mg\,{\sc i} line. \\
\citet{Bourrier2014} reported that the electron density needed to reproduce the detected Mg\,{\sc i} signature was surprisingly high. Here, acknowledging possible overestimation of the electron density (possibly due to erroneous radiative and dielectronic recombination rates, higher temperature, or an incorrect estimate of the Mg\,{\sc ii} absorption), we assumed a collisionless exosphere and used no input on this quantity. The model of \citet{Bourrier2014} was used to constrain the dynamics (and not the ionization) of the magnesium atoms and predict the absorption profile at 2853\,\AA\ from scaling the detection in HD\,209458b. Hereafter parameters labeled 209 refer to the planet HD\,209458b. \\
Physical properties of exoplanets were extracted from online databases (\citealt{Wright2011}; \citealt{Schneider2011}). In September 2013, 142 transiting planets had all the parameters required by the model, in particular to calculate the escape rate of neutral magnesium with the energy budget (e.g., \citealt{Ehrenreich_desert2011}) and to estimate the Mg\,{\sc i} line spectrum of each host star.\\


\noindent
\textit{Estimation of the planetary system properties}\\

Neutral magnesium atoms escape the atmosphere of a planet at the exobase radius $R_\mathrm{exo}$. It was fixed as a best guess by scaling the atmospheric structure of the planet to that of HD\,209458b, using the atmospheric scale height $H$, except when the resulting altitude is higher than the Roche lobe,
\begin{equation}
\label{eq:rlaunch}
R_\mathrm{exo}=min\left(R_\mathrm{p}+\frac{H}{H^\mathrm{209}} \left(R_\mathrm{exo}^\mathrm{209}-R_\mathrm{p}^\mathrm{209}\right)\, , \, R_\mathrm{Roche} \right).                          
\end{equation}

Atmospheric mass loss depends on the X/EUV stellar energy flux received by the upper atmosphere. We calculated the potential energy with Eq. (12) in \citet{Erkaev2007}, which takes into account the contribution of tidal forces through a correction factor $K_\mathrm{tide}$. The escape rate of neutral magnesium was obtained using the atmospheric mass fraction of this species, $A_\mathrm{Mg}$, so that
\begin{equation}
\label{eq:Mg_esc_ratebis}
\dot{M}_\mathrm{Mg}= A_\mathrm{Mg} \, \eta \, \frac{3 \, F_\mathrm{X/EUV}(\mathrm{1 AU})}{4 \, G \, d_\mathrm{tr}^{2} \, \overline{\rho} \, K_{tide}},   
\end{equation}
with $\overline{\rho}$ the mean density of the planet, $F_\mathrm{X/EUV}(\mathrm{1 AU})$ the X/EUV flux per unit area at 1\,AU from the star, $d_\mathrm{tr}$ the star-to-planet distance at the center of the transit, and $\eta$ the heating efficiency. We assumed that giant gaseous planets with $\overline{\rho}<$3\,g\,cm$^{-3}$ have the chemical composition of the Sun ($A_\mathrm{Mg}$=7.08$\times10^{-4}$; \citealt{Asplund2009}). For the 34 denser planets in our list of transiting planets we used the mass fraction of neutral magnesium found in the meteorites of the solar system ($A_\mathrm{Mg}$=9.55$\times10^{-2}$). For 55\,Cnc, HD\,209458, HD\,189733, and GJ\,436 we used the X/EUV stellar emission determined by \citet{Sanz-Forcada2011}. For 129 other stars we calculated the stellar EUV flux from its correlation with the rotation velocity of the star (e.g., \citealt{Wood1994}; \citealt{Ehrenreich_desert2011}). For the 9 remaining stars we estimated the EUV luminosity as a function of the stellar type using Eq. (1) in \citet{Lecav2007}. The heating efficiency $\eta$ is one of the less constrained values of the physics of exoplanet atmospheres ({\it e.g.}, \citealt{Lammer2009a}; \citealt{Owen2012}). We calculated escape rates in two limit scenarii, either with $\eta$=30\%, or with $\eta$=100\%. We fixed 30\% as a lower limit because it yields a 1$\sigma$ detection in the range -62 to -19\,km\,s$^{-1}$ for HD\,209458b, while the absorption signature was detected at 3.6$\sigma$.\\

For the Mg\,{\sc i} line stellar profiles we used the spectrum of HD\,209458b. This spectrum is scaled to a given star with the coefficient $\alpha_\mathrm{flux}$ , which takes into account the stellar type and the distance from Earth to the star,
\begin{align}
\label{eq:flux_stis_ref}
\alpha_{flux}=&\frac{F^\mathrm{Mg\,{\sc I}}_{*}}{F^\mathrm{Mg\,{\sc I}}_{209}} \\
                                                 =&10^{-0.4\,(V_{*}- V_{\mathrm{209}})-1.2\,((B-V)_{*} - (B-V)_{\mathrm{209}})}.     \nonumber       
\end{align}
This scaling law is derived from measurements taken from the IUE archive. For 55 Cnc we used the IUE spectrum of the G8V star Tau Ceti. For WASP-33 we used the emission spectrum of $\beta$ Pictoris (A6 star), retrieved from the STScI archive.\\
The dynamics of particles escaping an exoplanet atmosphere is naturally dependent on radiation pressure from the host star. For a magnesium atom the ratio of the radiation force to the stellar gravity $\beta_\mathrm{*}$ is proportional to the stellar Mg\,{\sc i} line flux received by the escaping atom at its radial velocity $v,$
\begin{equation}
\label{eq:beta}
\beta_{*}(\mathrm{v})=c^{\mathrm{Mg\,{\sc I}}}\, M_{*}^{-1} \, F^\mathrm{Mg\,{\sc I}}_{*}(1\,\mathrm{AU},\mathrm{v})   
,\end{equation}
with $c^{\mathrm{Mg\,{\sc I}}}$ a coefficient that only depends on the properties of the Mg\,{\sc I} line and the atom of magnesium, $M_{*}$ the mass of the star, and $F^\mathrm{Mg\,{\sc I}}_{*}$ the stellar flux at 1\,AU from the star at the wavelength corresponding to the Doppler velocity $v$. \\

The spectral errors on the fluxes were estimated assuming that they are dominated by photon noise (i.e., proportional to the square root of the flux, using the gain of real STIS observations of HD\,209458), and then propagated to estimate the uncertainty on the absorption signal. 

\section{Detectability of escaping magnesium particles}

\begin{figure}
\includegraphics[trim=2.3cm 2.5cm 4cm 3cm, clip=true,width=\columnwidth]{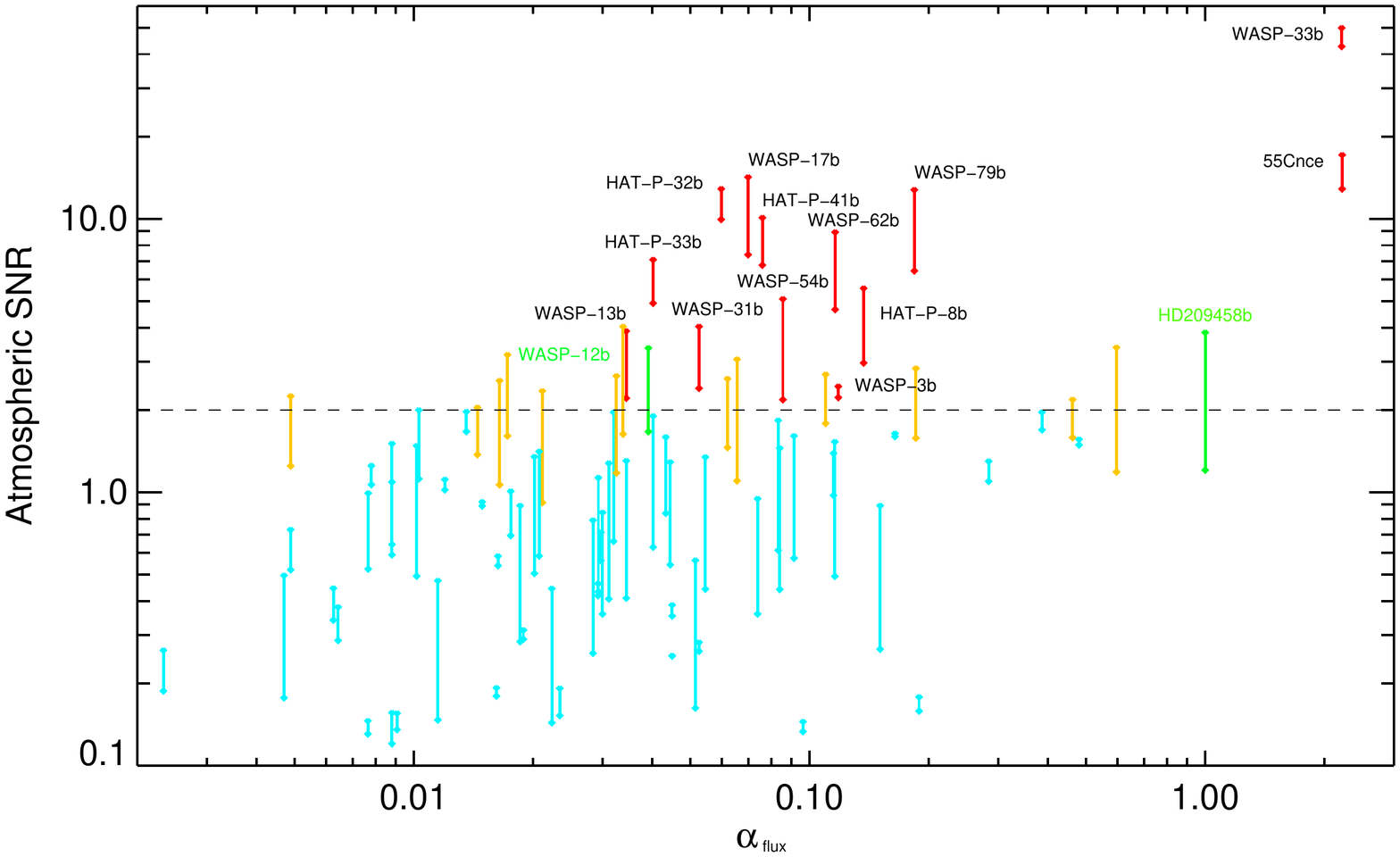}
\caption[]{S/N of absorption signatures generated by neutral magnesium in the atmospheres of known transiting exoplanets as a function of the brightness of their host stars in the Mg\,{\sc i} line seen from Earth. For each planet the lower (upper) S/N corresponds to a heating efficiency $\eta$=30\% (100\%). Planets with an S/N greater than 2 for $\eta=$30\% are plotted in red, and planets with an S/N greater than 2 for higher heating efficiencies in orange. Other planets are plotted in blue and the specific hot Jupiters HD\,209458 b and WASP-12 b
in green.}
\label{SN_targets_ffact}
\end{figure}

\begin{figure}
\includegraphics[trim=1.3cm 2cm 3cm 2cm, clip=true,width=\columnwidth]{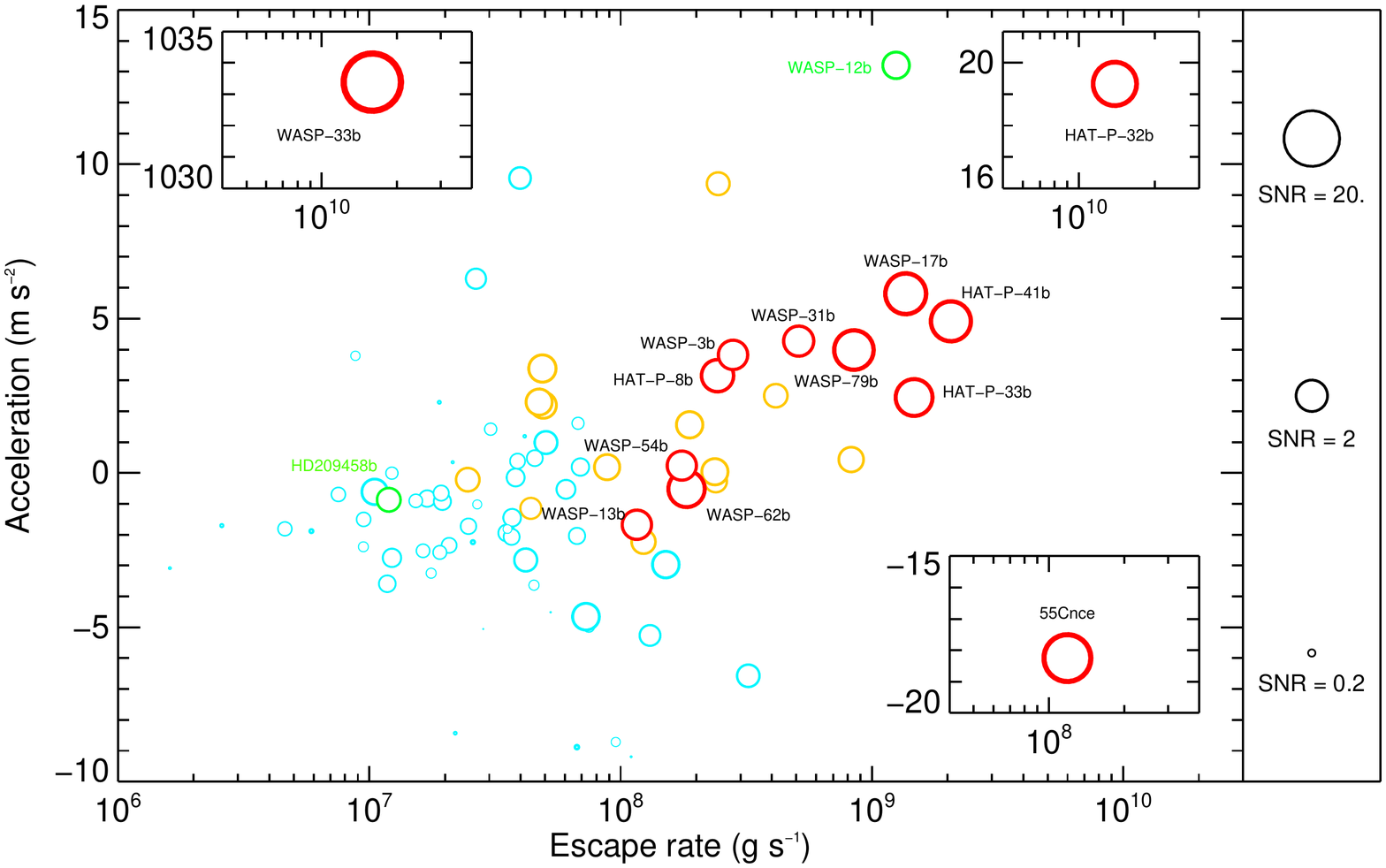}
\caption[]{S/N as a function of the escape rate and the radiation-pressure-induced acceleration on escaping particles. Each planet is represented by a disk whose size varies with the S/N (heating efficiency is 30\%). Colors are the same as in Fig.~\ref{SN_targets_ffact}. Most planets with a high S/N are found to have high escape rates and high accelerations.} 
\label{tau_force}       
\end{figure}

Magnesium particles escaping an exoplanet atmosphere are detected through their absorption profile in the Mg\,{\sc i} line of the host star, and we thus studied the signal-to-noise ratio (S/N) of this absorption signature. In Fig.~\ref{SN_targets_ffact} we show the S/N for the 142 exoplanets we studied, as a function of $\alpha_\mathrm{flux}$ for heating efficiencies between 30\% and 100\%. We caution that the S/N were calculated as if the planet transits were observed with HST/STIS in the same conditions as HD\,209458b, that is, for a total of about eight hours of observations with the instrument efficiency at 2853\,\AA. Thirteen planets stand out because of the high detectability of their extended atmospheres: \mbox{55\,Cnc e}, \mbox{HAT-P-8 b}, \mbox{HAT-P-33 b}, \mbox{HAT-P-41 b}, \mbox{WASP-3 b}, \mbox{WASP-13 b}, \mbox{WASP-17 b}, \mbox{WASP-31 b}, \mbox{WASP-33 b}, \mbox{WASP-54 b}, \mbox{WASP-62 b}, \mbox{WASP-79 b}, and \mbox{HAT-P-32 b}. Their evaporation state is such that they would still yield an S/N between 2 and 40 even with a conservative heating efficiency of 30\% (Table \ref{table_pl}). We caution that the evaporation of WASP-33 b and HAT-P-32 b is most likely limited by a low heating efficiency becaue with $\eta$=100\% these planets would evaporate in 1.6\,Gy and 900\,My, while they orbit stars aged 10-400\,Myr (\citealt{Cameron2010}; \citealt{Moya2011}) and 3.8$\stackrel{+1.5}{_{-0.5}}$Gy (\citealt{hartman2011}). There are thirteen other planets whose evaporating atmosphere would be detectable if they had high heating efficiencies (S/N$>$2 with $\eta$=100$\%$): \mbox{HAT-P-6 b}, \mbox{HAT-P-9 b}, \mbox{HAT-P-24 b}, \mbox{HAT-P-40 b}, \mbox{HD\,149026 b}, \mbox{HD\,209458 b}, \mbox{KELT-2A b}, \mbox{KELT-3 b}, \mbox{Kepler-20 b}, \mbox{TrES-4 b}, \mbox{WASP-1 b}, \mbox{WASP-12 b}, \mbox{WASP-67 b}, \mbox{WASP-78 b}, and \mbox{XO-4 b}. One of them is HD\,209458b, for which a positive detection of escaping magnesium was obtained, and WASP-12b, for which ionized magnesium was detected at high altitude in the exosphere (\citealt{Fossati2010}; \citealt{Haswell2012}). A higher heating efficiency may not necessarily lead to a proportionally higher S/N because the absorption depth depends not only on the escape rate, but also on self-shielding effects and the optical thickness of the extended atmosphere. In Fig.~\ref{tau_force} we display the S/N as a function of the escape rate and the radiation-pressure-induced acceleration of escaping particles, which control the absorption depth and the spectral range of the absorption signature. The combination of high escape rates with high acceleration generates the most significant absorption signatures, which is favored by planets orbiting bright stars at short orbital distances. Interestingly, the hot Jupiter HD\,189733 b, usually one of the best targets for atmospheric observations (in the Lyman-$\alpha$ line and from the near-UV to the mid-IR), is a poor candidate for observations of magnesium escape. This is mainly due to the low flux in the Mg\,{\sc i} line of its K2V star.

\setlength{\tabcolsep}{3.5pt}

\begin{table}
\caption[]{Properties of the planets of interest and their host stars. S/N are given for $\eta$=30\% and 100\%. Escape rates are calculated for neutral magnesium ($\dot{M}_\mathrm{Mg}$) and the total material in the atmosphere ($\dot{M}_\mathrm{t}$) when $\eta$=30\%, while $\dot{M}_\mathrm{t3\sigma}$ indicates the total escape rate required to obtain an S/N of 3.}
\label{table_pl}
\begin{tabular}{lccccc}
\hline

\noalign{\smallskip}  
Planet    & $M_\mathrm{p}$    &  log$\dot{M}_\mathrm{Mg}$       &        log$\dot{M}_\mathrm{t}$                  & log$\dot{M}_\mathrm{t3\sigma}$  & S/N                       \\         
          &($M_\mathrm{jup}$)   &        (g\,s$^{-1}$)                              &     (g\,s$^{-1}$)                                           &        (g\,s$^{-1}$)                             &  \\    
\noalign{\smallskip}
\hline
\noalign{\smallskip}
WASP-33 b       & 2.05 & 10.20  & 13.35 & 11.04         & 42.7/49.9\\

55\,Cnc e       & 0.03 & 8.07   & 9.09  &  7.82         & 12.9/17.1\\

HAT-P-32 b      & 0.95 & 10.14  & 13.29 &  12.25        & 9.9/12.9\\

WASP-17 b       & 0.51 & 9.13   & 12.28 &   11.50       & 7.4/14.2\\

HAT-P-41 b      & 0.80 & 9.31   & 12.46 &  11.75        & 6.8/10.1\\

WASP-79 b       & 0.89 & 8.93   & 12.08 & 11.42                 & 6.4/12.8\\

HAT-P-33 b      & 0.77 & 9.17   & 12.32 &  11.89        & 4.9/7.1\\

WASP-62 b       & 0.56 & 8.26   & 11.41 & 11.02                 & 4.7/8.9\\

HAT-P-8 b       & 1.29 & 8.39   & 11.54 &  11.54        & 3.0/5.6\\

WASP-31 b       & 0.48 & 8.71   & 11.86 & 12.05                 & 2.4/4.0\\

WASP-3 b                & 2.01 & 8.45   & 11.60 &    11.87  & 2.2/2.4\\

WASP-13 b       & 0.48 & 8.06   & 11.21 & 11.48                 & 2.2/3.9\\

WASP-54 b       & 0.63 & 8.24   & 11.39 &  11.67        & 2.2/5.1\\
\noalign{\smallskip}
\hline
\noalign{\smallskip}
WASP-12 b       & 1.36 & 9.10   & 12.25 &  12.74        & 1.7/3.4\\

HD\,209458 b& 0.69 & 7.08       & 10.23 &  11.03        & 1.2/3.8\\

\noalign{\smallskip}  
\hline
\end{tabular}
\end{table}

\section{Discussion and conclusion}

Escaping atmospheres have been detected in a very limited number of exoplanets, mainly through transit observations of the H\,{\sc i} Lyman-$\alpha$ line, and more recently, of the Mg\,{\sc i} line. While neutral hydrogen allows characterizing the exosphere far from the planet, magnesium can be used in addition to probe lower altitudes in the transition region between the thermosphere and the exosphere. We used a 3D model of atmospheric escape, initially developed for neutral magnesium escaping HD\,209458 b (\citealt{VM2013}; \citealt{Bourrier2014}), to simulate the escape of this heavy species from the atmospheres of known transiting exoplanets and to evaluate the resulting absorption signatures in the Mg\,{\sc i} line. The detectability of such signatures depends on the escape rate of neutral magnesium, the brightness and size of the star, and on the strength of its radiation pressure. We identified a sample of thirteen planets (55\,Cnc e, HAT-P-8 b, HAT-P-33 b, HAT-P-41 b, WASP-3 b, WASP-13 b, WASP-17 b, WASP-31 b, WASP-33 b, WASP-54 b, WASP-62 b, WASP-79 b, and HAT-P-32 b) with high escape rates of ten to a thousand times higher than for HD\,209458b, which are expected to produce atmospheric signatures that can be detected with current UV facilities such as the HST (they have an S/N of between about 2 and 50 in the least favorable conditions). Because of the detection of ionized magnesium in its exosphere, WASP-12 b would also be an interesting planet to search for neutral magnesium. This sample covers a wide range of planetary and stellar properties, from the very-hot Earth-like planet 55 Cnc e to the young A-star orbiting planet WASP-33 b. Observations of this sample would allow comparing exoplanet upper atmospheres and provide an unprecedented vision of the blow-off mechanism with constraints on the velocity, temperature, and density structure of the escaping gas in the thermosphere
- exosphere transition region (\citealt{Bourrier2014}). We can also anticipate that observations in the UV with the HST/STIS instrument would not only allow detecting neutral and ionized magnesium in their atmospheres, but also very likely the detection of many other heavy species such as Mn\,{\sc i}, Mn\,{\sc ii}, Fe\,{\sc i,} or Fe\,{\sc ii}.\\


\begin{acknowledgements}
We thank the referee, and offer particular thanks to the editor for its support. The authors acknowledge the support of the French Agence Nationale de la Recherche (ANR), under program ANR-12-BS05-0012 "Exo-Atmos". This work has also been supported by an award from the Fondation Simone et Cino Del Duca. This research has made use of the Extrasolar Planets Encyclopaedia at exoplanet.eu as well as the Exoplanet Orbit Database and the Exoplanet Data Explorer at exoplanets.org.
\end{acknowledgements}

\bibliographystyle{aa} 
\bibliography{biblio} 

\begin{thebibliography}{27}
\expandafter\ifx\csname natexlab\endcsname\relax\def\natexlab#1{#1}\fi

\bibitem[{{Asplund} {et~al.}(2009){Asplund}, {Grevesse}, {Sauval}, \&
  {Scott}}]{Asplund2009}
{Asplund}, M., {Grevesse}, N., {Sauval}, A.~J., \& {Scott}, P. 2009, \araa, 47,
  481

\bibitem[{{Ben-Jaffel} \& {Sona Hosseini}(2010)}]{BJ_Hosseini2010}
{Ben-Jaffel}, L. \& {Sona Hosseini}, S. 2010, \apj, 709, 1284

\bibitem[{{Bourrier} \& {Lecavelier des Etangs}(2013)}]{Bourrier_lecav2013}
{Bourrier}, V. \& {Lecavelier des Etangs}, A. 2013, \aap, 557, A124

\bibitem[{{Bourrier} {et~al.}(2014){Bourrier}, {Lecavelier des Etangs}, \&
  {Vidal-Madjar}}]{Bourrier2014}
{Bourrier}, V., {Lecavelier des Etangs}, A., \& {Vidal-Madjar}, A. 2014, \aap,
  565, A105

\bibitem[{{Collier Cameron} {et~al.}(2010){Collier Cameron}, {Guenther},
  {Smalley}, {McDonald}, {Hebb}, {Andersen}, {Augusteijn}, {Barros}, {Brown},
  {Cochran}, {Endl}, {Fossey}, {Hartmann}, {Maxted}, {Pollacco}, {Skillen},
  {Telting}, {Waldmann}, \& {West}}]{Cameron2010}
{Collier Cameron}, A., {Guenther}, E., {Smalley}, B., {et~al.} 2010, \mnras,
  407, 507

\bibitem[{{Ehrenreich} {et~al.}(2012){Ehrenreich}, {Bourrier}, {Bonfils},
  {Lecavelier des Etangs}, {H{\'e}brard}, {Sing}, {Wheatley}, {Vidal-Madjar},
  {Delfosse}, {Udry}, {Forveille}, \& {Moutou}}]{Ehrenreich2012}
{Ehrenreich}, D., {Bourrier}, V., {Bonfils}, X., {et~al.} 2012, \aap, 547, A18

\bibitem[{{Ehrenreich} \& {D{\'e}sert}(2011)}]{Ehrenreich_desert2011}
{Ehrenreich}, D. \& {D{\'e}sert}, J.-M. 2011, \aap, 529, A136

\bibitem[{{Erkaev} {et~al.}(2007){Erkaev}, {Kulikov}, {Lammer}, {Selsis},
  {Langmayr}, {Jaritz}, \& {Biernat}}]{Erkaev2007}
{Erkaev}, N.~V., {Kulikov}, Y.~N., {Lammer}, H., {et~al.} 2007, \aap, 472, 329

\bibitem[{{Fossati} {et~al.}(2010){Fossati}, {Haswell}, {Froning}, {Hebb},
  {Holmes}, {Kolb}, {Helling}, {Carter}, {Wheatley}, {Collier Cameron},
  {Loeillet}, {Pollacco}, {Street}, {Stempels}, {Simpson}, {Udry}, {Joshi},
  {West}, {Skillen}, \& {Wilson}}]{Fossati2010}
{Fossati}, L., {Haswell}, C.~A., {Froning}, C.~S., {et~al.} 2010, \apjl, 714,
  L222

\bibitem[{{Hartman} {et~al.}(2011){Hartman}, {Bakos}, {Torres}, {Latham},
  {Kov{\'a}cs}, {B{\'e}ky}, {Quinn}, {Mazeh}, {Shporer}, {Marcy}, {Howard},
  {Fischer}, {Johnson}, {Esquerdo}, {Noyes}, {Sasselov}, {Stefanik},
  {Fernandez}, {Szklen{\'a}r}, {L{\'a}z{\'a}r}, {Papp}, \&
  {S{\'a}ri}}]{hartman2011}
{Hartman}, J.~D., {Bakos}, G.~{\'A}., {Torres}, G., {et~al.} 2011, \apj, 742,
  59

\bibitem[{{Haswell} {et~al.}(2012){Haswell}, {Fossati}, {Ayres}, {France},
  {Froning}, {Holmes}, {Kolb}, {Busuttil}, {Street}, {Hebb}, {Collier Cameron},
  {Enoch}, {Burwitz}, {Rodriguez}, {West}, {Pollacco}, {Wheatley}, \&
  {Carter}}]{Haswell2012}
{Haswell}, C.~A., {Fossati}, L., {Ayres}, T., {et~al.} 2012, \apj, 760, 79

\bibitem[{{Kulow} {et~al.}(2014){Kulow}, {France}, {Linsky}, \&
  {Loyd}}]{Kulow2014}
{Kulow}, J.~R., {France}, K., {Linsky}, J., \& {Loyd}, R.~O.~P. 2014, \apj,
  786, 132

\bibitem[{{Lammer} {et~al.}(2009){Lammer}, {Odert}, {Leitzinger},
  {Khodachenko}, {Panchenko}, {Kulikov}, {Zhang}, {Lichtenegger}, {Erkaev},
  {Wuchterl}, {Micela}, {Penz}, {Biernat}, {Weingrill}, {Steller}, {Ottacher},
  {Hasiba}, \& {Hanslmeier}}]{Lammer2009a}
{Lammer}, H., {Odert}, P., {Leitzinger}, M., {et~al.} 2009, \aap, 506, 399

\bibitem[{{Lecavelier des Etangs}(2007)}]{Lecav2007}
{Lecavelier des Etangs}, A. 2007, \aap, 461, 1185

\bibitem[{{Lecavelier des Etangs} {et~al.}(2010){Lecavelier des Etangs},
  {Ehrenreich}, {Vidal-Madjar}, {Ballester}, {D{\'e}sert}, {Ferlet},
  {H{\'e}brard}, {Sing}, {Tchakoumegni}, \& {Udry}}]{Lecav2010}
{Lecavelier des Etangs}, A., {Ehrenreich}, D., {Vidal-Madjar}, A., {et~al.}
  2010, \aap, 514, A72

\bibitem[{{Lecavelier des Etangs} {et~al.}(2004){Lecavelier des Etangs},
  {Vidal-Madjar}, {McConnell}, \& {H{\'e}brard}}]{Lecav2004}
{Lecavelier des Etangs}, A., {Vidal-Madjar}, A., {McConnell}, J.~C., \&
  {H{\'e}brard}, G. 2004, \aap, 418, L1

\bibitem[{{Linsky} {et~al.}(2010){Linsky}, {Yang}, {France}, {Froning},
  {Green}, {Stocke}, \& {Osterman}}]{Linsky2010}
{Linsky}, J.~L., {Yang}, H., {France}, K., {et~al.} 2010, \apj, 717, 1291

\bibitem[{{Moya} {et~al.}(2011){Moya}, {Bouy}, {Marchis}, {Vicente}, \&
  {Barrado}}]{Moya2011}
{Moya}, A., {Bouy}, H., {Marchis}, F., {Vicente}, B., \& {Barrado}, D. 2011,
  \aap, 535, A110

\bibitem[{{Owen} \& {Jackson}(2012)}]{Owen2012}
{Owen}, J.~E. \& {Jackson}, A.~P. 2012, \mnras, 425, 2931

\bibitem[{{Sanz-Forcada} {et~al.}(2011){Sanz-Forcada}, {Micela}, {Ribas},
  {Pollock}, {Eiroa}, {Velasco}, {Solano}, \&
  {Garc{\'{\i}}a-{\'A}lvarez}}]{Sanz-Forcada2011}
{Sanz-Forcada}, J., {Micela}, G., {Ribas}, I., {et~al.} 2011, \aap, 532, A6

\bibitem[{{Schlawin} {et~al.}(2010){Schlawin}, {Agol}, {Walkowicz}, {Covey}, \&
  {Lloyd}}]{Schlawin2010}
{Schlawin}, E., {Agol}, E., {Walkowicz}, L.~M., {Covey}, K., \& {Lloyd}, J.~P.
  2010, \apjl, 722, L75

\bibitem[{{Schneider} {et~al.}(2011){Schneider}, {Dedieu}, {Le Sidaner},
  {Savalle}, \& {Zolotukhin}}]{Schneider2011}
{Schneider}, J., {Dedieu}, C., {Le Sidaner}, P., {Savalle}, R., \&
  {Zolotukhin}, I. 2011, \aap, 532, A79

\bibitem[{{Vidal-Madjar} {et~al.}(2004){Vidal-Madjar}, {D{\'e}sert},
  {Lecavelier des Etangs}, {H{\'e}brard}, {Ballester}, {Ehrenreich}, {Ferlet},
  {McConnell}, {Mayor}, \& {Parkinson}}]{VM2004}
{Vidal-Madjar}, A., {D{\'e}sert}, J.-M., {Lecavelier des Etangs}, A., {et~al.}
  2004, \apjl, 604, L69

\bibitem[{{Vidal-Madjar} {et~al.}(2013){Vidal-Madjar}, {Huitson}, {Bourrier},
  {D{\'e}sert}, {Ballester}, {Lecavelier des Etangs}, {Sing}, {Ehrenreich},
  {Ferlet}, {H{\'e}brard}, \& {McConnell}}]{VM2013}
{Vidal-Madjar}, A., {Huitson}, C.~M., {Bourrier}, V., {et~al.} 2013, \aap, 560,
  A54

\bibitem[{{Vidal-Madjar} {et~al.}(2003){Vidal-Madjar}, {Lecavelier des Etangs},
  {D{\'e}sert}, {Ballester}, {Ferlet}, {H{\'e}brard}, \& {Mayor}}]{VM2003}
{Vidal-Madjar}, A., {Lecavelier des Etangs}, A., {D{\'e}sert}, J.-M., {et~al.}
  2003, \nat, 422, 143

\bibitem[{{Wood} {et~al.}(1994){Wood}, {Brown}, {Linsky}, {Kellett}, {Bromage},
  {Hodgkin}, \& {Pye}}]{Wood1994}
{Wood}, B.~E., {Brown}, A., {Linsky}, J.~L., {et~al.} 1994, \apjs, 93, 287

\bibitem[{{Wright} {et~al.}(2011){Wright}, {Fakhouri}, {Marcy}, {Han}, {Feng},
  {Johnson}, {Howard}, {Fischer}, {Valenti}, {Anderson}, \&
  {Piskunov}}]{Wright2011}
{Wright}, J.~T., {Fakhouri}, O., {Marcy}, G.~W., {et~al.} 2011, \pasp, 123, 412

\end{thebibliography}

\end{document}